\documentclass[twocolumn]{aastex631}

\usepackage{tikz}
\usetikzlibrary{matrix}
\usepackage[dvipsnames]{xcolor}
\usepackage{amsmath}

\begin{document}

\title{Linking Warm Dark Matter to Merger Tree Histories via Deep Learning Networks}

\correspondingauthor{Ilem Leisher} \\
\email{ilemleisher@gmail.com}

\author[0009-0006-5740-5318]{Ilem Leisher}
\affiliation{Grinnell College, 1115 8th Ave, Grinnell, IA 50112, USA}
\affiliation{Department of Physics, Massachusetts Institute of Technology, Cambridge, MA 02139, USA}
\affiliation{Kavli Institute for Astrophysics and Space Research, Massachusetts Institute of Technology, Cambridge, MA 02139, USA}

\author[0000-0002-5653-0786]{Paul Torrey}
\affiliation{Department of Astronomy, University of Virginia, 530 McCormick Road, Charlottesville, VA 22904 USA}
\affiliation{The NSF-Simons AI Institute for Cosmic Origins, USA}
\affiliation{Virginia Institute for Theoretical Astronomy, University of Virginia, Charlottesville, VA 22904, USA}

\author[0000-0002-8111-9884]{Alex M. Garcia}
\affiliation{Department of Astronomy, University of Virginia, 530 McCormick Road, Charlottesville, VA 22904 USA}
\affiliation{The NSF-Simons AI Institute for Cosmic Origins, USA}
\affiliation{Virginia Institute for Theoretical Astronomy, University of Virginia, Charlottesville, VA 22904, USA}

\author[0000-0002-2628-0237]{Jonah C. Rose}
\affiliation{Center for Computational Astrophysics, Flatiron Institute, 162 5th Avenue, New York, NY 10010, USA}
\affiliation{Department of Physics, Princeton University, Princeton, NJ 08544, USA}

\author[0000-0002-4816-0455]{Francisco Villaescusa-Navarro}
\affiliation{Center for Computational Astrophysics, Flatiron Institute, 162 5th Avenue, New York, NY 10010, USA}
\affiliation{Department of Physics, Princeton University, Princeton, NJ 08544, USA}

\author[0000-0001-9543-9910]{Zachary Lubberts}
\affiliation{Department of Statistics, University of Virginia, Charlottesville, VA}

\author[0000-0003-0777-4618]{Arya~Farahi}
\affiliation{Departments of Statistics and Data Science, University of Texas at Austin, Austin, TX 78757, USA}
\affiliation{The NSF-Simons AI Institute for Cosmic Origins, University of Texas at Austin, Austin, TX 78712, USA}

\author[0000-0002-7968-2088]{Stephanie O'Neil}
\affiliation{Department of Physics \& Astronomy, University of Pennsylvania, Philadelphia, PA 19104, USA}
\affiliation{Department of Physics, Princeton University, Princeton, NJ 08544, USA}

\author[0000-0002-6196-823X]{Xuejian Shen}
\affiliation{Kavli Institute for Astrophysics and Space Research, Massachusetts Institute of Technology, Cambridge, MA 02139, USA}

\author[0009-0009-0239-8706]{Olivia Mostow}
\affil{Department of Astronomy, University of Virginia, 530 McCormick Road, Charlottesville, VA 22904 USA}

\author[0000-0002-3204-1742]{Nitya Kallivayalil}
\affiliation{Department of Astronomy, University of Virginia, 530 McCormick Road, Charlottesville, VA 22904 USA}
\affiliation{The NSF-Simons AI Institute for Cosmic Origins, USA}

\author[0009-0008-7017-5742]{Dhruv Zimmerman}
\affiliation{Department of Astronomy, University of Florida, 211 Bryant Space Sciences Center, Gainesville, FL 32611 USA}

\author[0000-0002-7064-4309]{Desika Narayanan}
\affiliation{Department of Astronomy, University of Florida, 211 Bryant Space Sciences Center, Gainesville, FL 32611 USA}
\affiliation{Cosmic Dawn Center at the Niels Bohr Institute, University of Copenhagen and DTU-Space, Technical University of Denmark}

\author[0000-0001-8593-7692]{Mark Vogelsberger}
\affiliation{Department of Physics, Massachusetts Institute of Technology, Cambridge, MA 02139, USA}
\affiliation{Kavli Institute for Astrophysics and Space Research, Massachusetts Institute of Technology, Cambridge, MA 02139, USA}
\affiliation{The NSF AI Institute for Artificial Intelligence and Fundamental Interactions, Cambridge, MA 02139, USA}
\affiliation{Fachbereich Physik, Philipps Universit\"at Marburg, D-35032 Marburg, Germany}

\begin{abstract}
Dark matter (DM) halos form hierarchically in the Universe through a series of merger events.
Cosmological simulations can represent this series of mergers as a graph-like ``tree'' structure.
Previous work has shown these merger trees are sensitive to cosmology simulation parameters, but as DM structures, the outstanding question of their sensitivity to DM models remains unanswered.
In this work, we investigate the feasibility of deep learning methods trained on merger trees to infer Warm Dark Matter (WDM) particles masses from the DREAMS simulation suite. 
We organize the merger trees from 1,024 zoom-in simulations into graphs with nodes representing the merger history of galaxies and edges denoting hereditary links. We vary the complexity of the node features included in the graphs ranging from a single node feature up through an array of several galactic properties (e.g., halo mass, star formation rate, etc.). We train a Graph Neural Network (GNN) to predict the WDM mass using the graph representation of the merger tree as input. We find that the GNN can predict the mass of the WDM particle ($R^2$ from 0.07 to 0.95), with success depending on the graph complexity and node features. 
We extend the same methods to supernovae and active galactic nuclei feedback parameters $A_\text{SN1}$, $A_\text{SN2}$, and $A_\text{AGN}$, successfully inferring the supernovae parameters.
The GNN can even infer the WDM mass from merger tree histories without any node features, indicating that the structure of merger trees alone inherits information about the cosmological parameters of the simulations from which they form. 

\end{abstract}


\section{Introduction} \label{sec:intro}

Within the standard $\Lambda$ cold dark matter ($\Lambda$CDM) model of cosmology, dark matter (DM) is assumed to be cold and collisionless. As a whole, CDM functions exceptionally well for the large-scale structures in the Universe \citep{2007ApJS..170..377S,deBlok_2008, 1991ApJ...378..496D,Dalal_2002,virmos}, but creates tensions at the sub-galactic level \citep{Nadler_2021,moore1994evidence,10.1111/j.1745-3933.2011.01074.x, papastergis2015,10.1093/mnras/stv1504,Klypin_1999, Moore_1999}.
These tensions can be alleviated via modified treatments of either the galaxy formation model, or modifications to the cosmological framework itself.

Due to the difficult nature of observing DM and its growth over time, cosmological simulations are extremely valuable to investigations of DM models. In particular, the recent use of Machine Learning (ML) for analyzing data and making inferences within these simulations has proven to be a powerful tool for the study of DM \citep{PhysRevD.104.123526, Villaescusa-Navarro_2022b}. For example, the CAMELS project \citep{Villaescusa-Navarro_2021, Villaescusa-Navarro_2023, Villaescusa-Navarro_2022} and the DREAMS project \citep{rose2025introducing} both employ ML methods in simulation suites tracking DM particles to study structure evolution as a function of cosmology and astrophysics. 

The traditional ML tool for this task has been Convolutional Neural Networks (CNNs), which have been successfully used to infer cosmological parameters \citep{10.1093/mnras/stad3260,villaescusanavarro2021multifieldcosmologyartificialintelligence,villaescusanavarro2021robustmarginalizationbaryoniceffects}. However, the capabilities of CNNs are limited by their inputs, which must occupy a linear, Euclidean space. This forces input data into a grid structure, overlooking non-linear spatial dependencies the data may have, and potentially resulting in a worse performance of the model. This has led to investigations of other ML methods, such as Graph Neural Networks \citep[GNNs;][]{4700287}, which offer architectures to capture complex non-linear spatial dependencies. GNNs learn from input data occupying graph spaces, which preserve such non-linear spatial relationships in the data. They have been shown to be an effective model for inferring both cosmological \citep{Villanueva-Domingo_2022} and astrophysical information \citep{jespersen2022mangrove, 2024arXiv241112629G, chuang2024leaving}, in some cases more so than CNNs \citep{jespersen2022mangrove}. 
Learning from non-linear spatial dependencies allow GNNs to extract more information from certain datasets than CNNs can.

There have been several examples of GNNs demonstrating their efficacy in inferring astrophysical and cosmological information within simulations \citep{Villanueva-Domingo_2022, 2024arXiv241112629G,beck2019refined, shao2023robust, de2023robust, de2023field, massara2023predicting,jespersen2022mangrove,chuang2024leaving}. Such capabilities have provided insight into how the evolution of particular structures (e.g., galaxy distributions, merger trees) are influenced by simulation parameters. For the study of DM, the examination of merger trees is particularly interesting, as these are structures that trace a DM halo's growth through merger events. Though recent work has successfully used GNNs to infer both cosmological parameters and galaxy properties from merger trees \citep{jespersen2022mangrove,chuang2024leaving,huang2025cosmobench}, it has so far not been shown if a GNN can infer DM properties (such as the particle mass) from merger trees. In this paper, we test this possibility by training a GNN on merger tree histories from cosmological simulations.

In this work we use a Warm Dark Matter (WDM) simulation suite from the DREAMS project \citep{rose2025introducing} to provide us with cosmological and astrophysical parameters for our GNN. 
A WDM model is a simple category of DM models, in that it has only one parameter, particle mass \citep{Bode_2001}, making it particularly compatible with ML methods \citep{10.1093/mnras/stad3260,Lin_2024} and thus fitting for our purposes. 
In contrast to CDM, a WDM model assumes somewhat lighter DM particles leading to higher streaming velocities in the early universe that wash out density fluctuations on small scales. 
Previous work has assessed whether WDM can accurately describe the DM features we observe in the Universe or not \citep{10.1111/j.1365-2966.2007.12053.x,lovell2012haloes,viel2013warm,lovell2014properties,fitts2019dwarf}. Primarily, the WDM particle mass has been constrained via comparisons to the observed satellite luminosity function in the Local Group \citep{kennedy2014constraining,murgia2017non,nadler2021constraints}, with the Lyman-alpha forest from high redshift quasars \citep{viel2013warm,irvsivc2017new}, and between small halo properties in strongly lensed systems \citep{ritondale2019low,10.1093/mnras/stz3480}. The strongest constraints have been placed on the particle mass using a combination of methods: \cite{Nadler_2021} and \cite{10.1093/mnras/stab1960} found upper and lower limits of 9.7 keV and 6.0 keV, respectively.

Although the model potentially alleviates CDM tensions such as the production of cored density profiles \citep{Bode_2001}, these constraints are incompatible with observational properties of dwarf galaxies, which require a WDM particle mass of $\sim2~{\rm keV}$ \citep{lovell2014properties,fitts2019dwarf}. Thus, WDM is an unlikely candidate for the entire DM population in the Universe \citep{10.1093/mnras/stab1960,Nadler_2021}. Regardless, demonstrating that a GNN can infer WDM properties from merger tree histories paves the way for doing so with more realistic models in the future, while also creating a framework where generative models for merger trees can appropriately reflect the cosmology from which they are drawn.
Following the work of \cite{Lin_2024}, we also infer three astrophysics parameters within the WDM suite that control supernovae (SN) and active galactic nuclei (AGN) feedback. This further tests the capabilities of our model as well as the information contained within the merger trees.

The structure of this paper is outlined as follows. In Section~\ref{sec:methods}, we describe the simulation suite from which we obtain our merger trees and our chosen GNN model. In Section~\ref{sec:results}, we present our results from training the GNNs on a variety of galaxy properties to infer the WDM and astrophysics parameters, focusing on the WDM parameter as it has the most interesting results. In Section~\ref{sec:discussion}, we discuss our results and investigate WDM sensitivities contained in the merger trees. Finally, in Section~\ref{sec:conclusions}, we state our conclusions.

\section{Methods}
\label{sec:methods}

Our goal is to build a framework to perform graph-based inference using GNNs that take the galaxy merger trees as input in order to predict cosmological and simulation parameters.  This requires (i) a simulation dataset with varied astrophysical and cosmological parameters, (ii) a procedure for converting merger trees into graph objects, and (iii) an appropriate GNN architecture.  We describe each of these elements in this section.

\subsection{Simulations} \label{sec:simulations}

Our data is from the WDM Milky Way zoom-in suite within the DaRk mattEr and Astrophysics with Machine learning and Simulations (DREAMS) project \citep{rose2025introducing}\ignorespaces
\footnote{\url{https://www.dreams-project.org/}}.
This suite is well-suited for our study as it has a broad range of WDM, cosmology, and astrophysics parameter variations while also maintaining high baryon mass resolution, which allows for reasonably detailed resolution of galaxy merger trees (sub)halo formation histories.
The suite contains 1,024 hydrodynamical simulation zoom-in boxes run using the IllustrisTNG galaxy formation model \citep{10.1093/mnras/stx2656,10.1093/mnras/sty1733}, which itself is based on the Illustris galaxy formation model~\citep{Vogelsberger2013, Torrey2014}, and advanced with the moving-mesh code {\sc Arepo} \citep{10.1111/j.1365-2966.2009.15715.x,2019ascl.soft09010S,Weinberger_2020}. 
Initial conditions are generated using {\sc Music} \citep{10.1111/j.1365-2966.2011.18820.x}, which evolves a uniform particle distribution through time using second-order Lagrangian perturbation equations. 
Each simulation spans $z=127$ to $z=0$ and includes 91 snapshots from $z=15$ to $z=0$. 
A (single) merger tree is created from each simulation using {\sc SubLink} \citep{10.1093/mnras/stv264} from subhalos identified with {\sc Subfind} \citep{10.1046/j.1365-8711.2001.04912.x}. The mass of subhalos in the suite ranges from $7 \times 10^{11}~\text{M}_\odot $ to $2.5 \times 10^{12} ~\text{M}_\odot $.
Compared to the TNG50-1 simulation, the highest resolution box of the IllustrisTNG suite \citep{10.1093/mnras/stz2306, 10.1093/mnras/stz2338}, the DREAMS zoom-in suites have $\sim5\times$ as many MW-mass galaxies at comparable baryon mass resolution (a factor of $\sim2$ lower). 

\subsubsection{Parameters} \label{sec:parameters}

Although the DREAMS suites are primarily based on the IllustrisTNG physics model, there are a number of systematic parameter variations of both the DM physics and astrophysics of the simulations. 
In this section, we detail the parameter variations within the WDM suite.

The simulations in this suite have fixed cosmological parameters that are consistent with \cite{refId0}: $\Omega_{\text{b}} = 0.046, h = 0.691, \Omega_{\text{m}} = 0.302, \Omega_{\Lambda} = 0.698, \sigma_8 = 0.839$.
In the suite we use, there are three varied astrophysical parameters, $A_{\text{SN1}}$, $A_{\text{SN2}}$, $A_{\text{AGN}}$, and one varied cosmological parameter, $M_\text{WDM}$. We now briefly explain each varied parameter (we refer the reader to \citeauthor{rose2025introducing} \citeyear{rose2025introducing} Section~2.1 for a complete description of the model).

The $A_{\text{SN1}}$ and $A_{\text{SN2}}$ parameters are dimensionless scaling factors in the IllustrisTNG SN feedback model. This model contains two main components, both of which depend on these scaling factors. The first component is the mass-loading factor $\eta_w$, which depends on the second component, stellar wind velocity $v_w$, such that $\eta_w \propto e_w/v_w^2$. $e_w$ is the specific energy available for generating winds, which scales based on $A_\text{SN1}$, according to the equation
\begin{multline}
    e_w = A_{\text{SN1}} \bar{e}_w \left[ f_{w,Z} + \frac{1 - f_{w,Z}}{1+ (Z/Z_{w,\text{ref}})^{\gamma_{w,Z}}} \right]\\ \times N_\text{SNII}~~E_\text{SNII,51}.
\end{multline}
In this equation, $Z$ is the metallicity of star-forming gas, $f_{w,Z}$ is included to reduce the available energy when the metallicity is above $Z_{w,\text{ref}}$, $\bar{e}_w$ is a dimensionless scaling factor set to $3.6$ in the fiducial TNG model, $N_\text{SNII}$ is the number of SN Type II per formed stellar mass, $E_\text{SNII,51}$ is the available energy per core collapse SNe, and $A_\text{SN1}$ is our varied scaling factor parameter. 

The second component, stellar wind velocity, is scaled by $A_\text{SN2}$, according to the equation
\begin{equation}
    v_w = \max \left[ A_{\text{SN2}}~ \kappa_w~\sigma_\text{DM} \left( \frac{H_0}{H(z)} \right)^{1/3}, v_{w,\text{min}} \right],
\end{equation}
where $H(z)$ is the Hubble parameter, $v_{w,\text{min}}$ is the minimum wind speed, $\kappa_w$ is a dimensionless scaling (set to $7.4$ in the fiducial model), $\sigma_\text{DM}$ is the local one-dimensional DM velocity dispersion, and $A_\text{SN2}$ is our varied scaling factor parameter. 

The $A_\text{AGN}$ parameter scales the fraction of energy transferred to nearby gases due to black hole accretion ($\epsilon_{f,\,{\rm high}}$). This is incorporated into the IllustrisTNG AGN feedback model via the feedback energy released from accreting black holes. Notably, black holes are not seeded in DM halos below $\sim7\times10^7 M_\odot$ \citep{10.1093/mnras/sty1733}. The equation for this feedback energy is 
\begin{equation}
    \label{eq:asn2}
    \Delta \dot{E} = A_\text{AGN} \epsilon_{f,\,{\rm high}} \epsilon_r {\dot{\text{M}}}_\text{BH}c^2,
\end{equation}
where $\epsilon_r$ is the radiative efficiency, which is the canonical 0.1–0.2 of the accreted rest-mass energy that is released in the accretion process, ${\dot{\text{M}}}_\text{BH}$ is the black hole accretion rate, and $A_\text{AGN}$ is our varied parameter.

The $M_\text{WDM}$ parameter is the WDM particle mass. WDM is implemented into the simulations following a suppression of the matter power spectrum from \cite{Bode_2001}:
\begin{equation}
    P_\text{WDM}(k) = \beta(k) P_\text{CDM}(k),
\end{equation}
where $P_\text{WDM}(k)$ and $P_\text{CDM}(k)$ are the WDM and CDM power spectra, respectively. $\beta(k)$ is given by
\begin{equation}
    \beta(k) = ((1+ (\alpha k)^{2.4})^{-5.0/1.2})^2~,
\end{equation}
and
\begin{multline}
    \alpha = 0.048 ~h^{-1}\left( M_\text{WDM} \right)^{-1.15}\\ \times\left( \frac{\Omega_{\rm m}-\Omega_{\rm b}}{0.4} \right)^{0.15} \left( \frac{h}{0.65} \right)^{1.3}
\end{multline}
where $\Omega_{\rm m}$ is the total matter density (fixed to 0.302 in this suite) and $\Omega_{\rm b}$ is the total baryon density.
The suppression factor models the free-streaming behavior of WDM particles, erasing small-scale power below a characteristic scale that is dependent on the assumed WDM particle mass.

These four parameters are varied following a Sobol sequence \citep{SOBOL196786} in order to uniformly sample across the parameter space.
Sobol sampling provides dense coverage of the specified parameter space while improving convergence rate compared to purely random sampling. The WDM suite uses the following parameter sampling ranges:

\begin{equation}
0.033 ~{\rm keV}^{-1} < 1/M_\text{WDM} < 0.555~{\rm keV}^{-1}
\end{equation}
\begin{equation}
0.25 < \log_{10}({A_\text{SN1}}) < 4
\end{equation}
\begin{equation}
0.5 < \log_{10}({A_\text{SN2}}) < 2
\end{equation}
\begin{equation}
0.25 < \log_{10}({A_\text{AGN}}) < 4
\end{equation}

\subsubsection{{\sc SubLink} Merger Trees}

The {\sc SubLink} merger trees are constructed by assigning a descendant to each subhalo in a given box based on a three-step process detailed in \cite{10.1093/mnras/stv264} and outlined here. First, for a given subhalo, all subhalos in the next snapshot backward in time that share common particles with the subhalo are tagged as candidates. Next, each candidate is given a score as a function of the binding energy rank of the common particles between the two particle-sharing subhalos. Finally, the candidate with the highest score is defined as the unique descendant of the given subhalo. 

Each subhalo in {\sc SubLink} trees is given pointers to five key subhalos \citep{springel2005large}. The first is the \textit{first progenitor}: the progenitor of a given subhalo with the most massive history. The second is the \textit{next progenitor}: the subhalo which shares the same descendant as a given subhalo, which has the next largest mass history. The third is the \textit{descendant}: the unique subhalo that both has common particles with a given subhalo and has the least binding energy rank summed across these particles. The fourth is the \textit{first subhalo in Friend-of-Friends group}: the main subhalo (defined as the one with the most massive history) from the Friend-of-Friends \citep[FoF;][]{Davis:1985rj} group of a given subhalo. The final key subhalo is the \textit{next subhalo in FoF group}: the next subhalo from the same FoF group, in order of decreasing mass history.

We walk each merger tree using a top-down approach moving backward in time, beginning with the MW subhalo at snapshot 90. We find this subhalo's first progenitor and all of its next progenitors, and then walk down the subtrees stemming from each of these subhalos, using a recursive function to repeat this action for every subhalo in the tree. Unlike previous work that has used trees with certain subhalos excluded \citep[see Section~\ref{sec:prune} for details;][]{jespersen2022mangrove,chuang2024leaving}, we elect to include all subhalos in our trees. Throughout this process, we select data properties from both the available {\sc SubLink} fields and the FoF group catalog associated with each subhalo, and save them in a manner that preserves the structure of the merger tree within a storage file.

\subsection{Graph Neural Network}

Graph representation of the merger trees preserves the galaxy assembly history information in a form that can be naturally ingested into GNNs.
In this section, we describe the specific form of our merger tree-based graphs, the architecture and application of our GNN model, as well as the loss function we use. 

\subsubsection{Graphs}

\begin{figure}
    \centering
    \includegraphics[width=\linewidth,trim={0cm 2cm 0cm 1.5cm},clip]{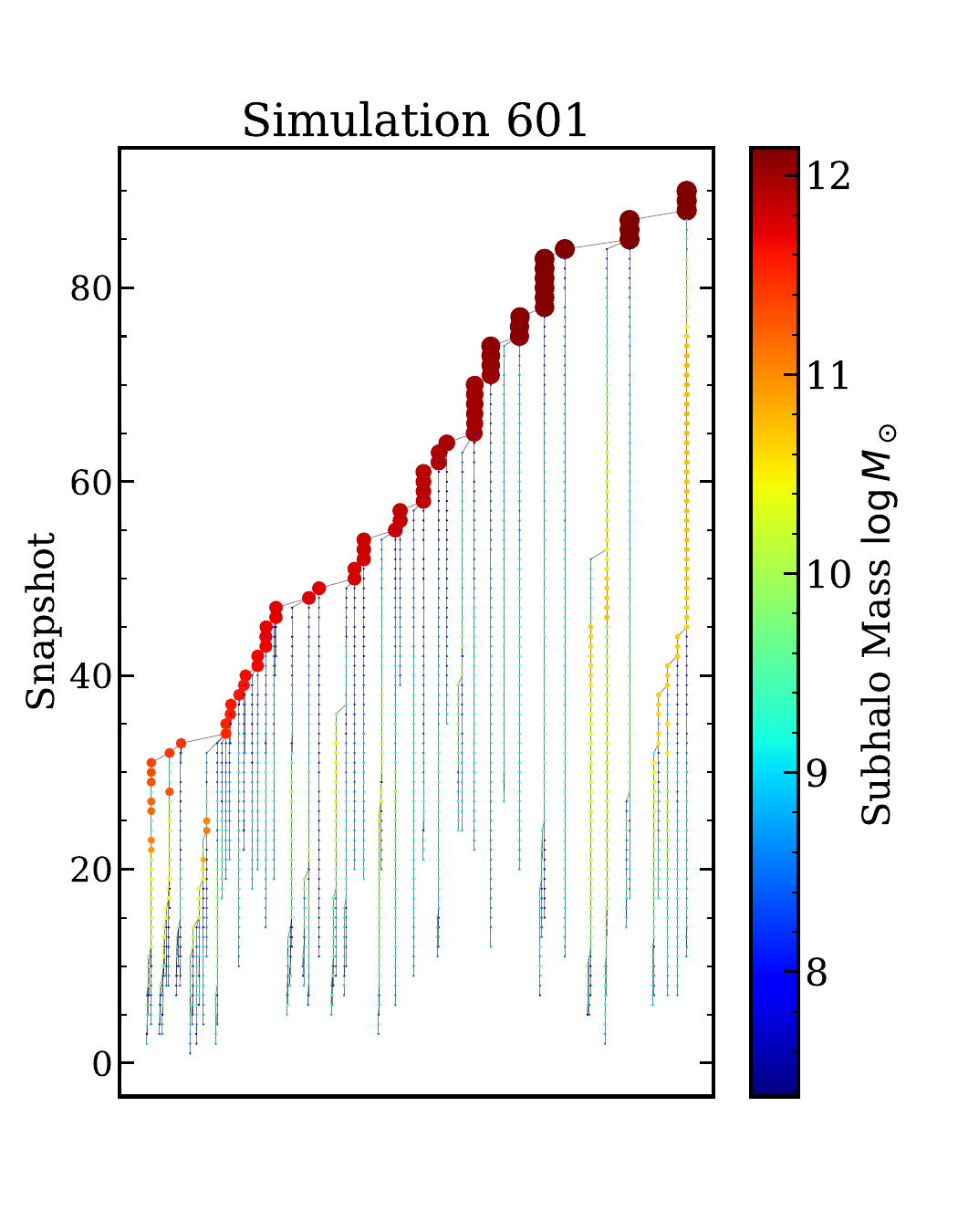}
    \caption{The graph representation of a merger tree from an example simulation within the DREAMS WDM MW zoom-in suite. The merger tree represents the merger history of a MW-mass subhalo beginning at $z=0$ (snapshot 90) and ending at $z=15$ (snapshot 0). All subhalos in the tree are included. The colors of the nodes, as well as the sizes of the nodes, represent the subhalo mass (ranging from $7 \times 10^{11}~\text{M}_\odot $ to $2.5 \times 10^{12} ~\text{M}_\odot $).}
    \label{fig:tree}
\end{figure}

We convert {\sc SubLink} merger trees into graph objects using the PyTorch Geometric package \citep{fey2019fastgraphrepresentationlearning}, visualized in Figure~\ref{fig:tree}. In this package, graph objects $G = (X, (I, E))$ are comprised of a node feature matrix $X$ containing node features $x$, and a sparse adjacency tuple $(I, E)$ containing edges directed backwards in time with edge indices encoded in $I$ and optional edge features for each edge in $E$. We map each data property in a given storage file to a node feature in the corresponding graph, and we create pairs of edge indices by matching each subhalo with its unique descendant to satisfy the edge tuple. We do not include any edge features, as these represent information contained within the edges between two nodes (i.e., spatial distance between two nodes), whereas we are only interested in the existence of an edge between two nodes. Though this choice nominally translates to a loss of information, we believe it is fitting for our methods of investigating merger trees. For the cosmological and astrophysical parameter values $M_{\text{WDM}}$, $A_\text{SN1}$, $A_\text{SN2}$, $A_\text{AGN}$ in each box, we normalize each value with $(A-\min(A))/(\max(A)-\min(A))$, where $A$ is a parameter and $\min(A)$ and $\max(A)$ are the bounds of the corresponding sampling space. We attach these normalized values to each associated graph.

\begin{figure}
    \centering
    \includegraphics[width=\linewidth]{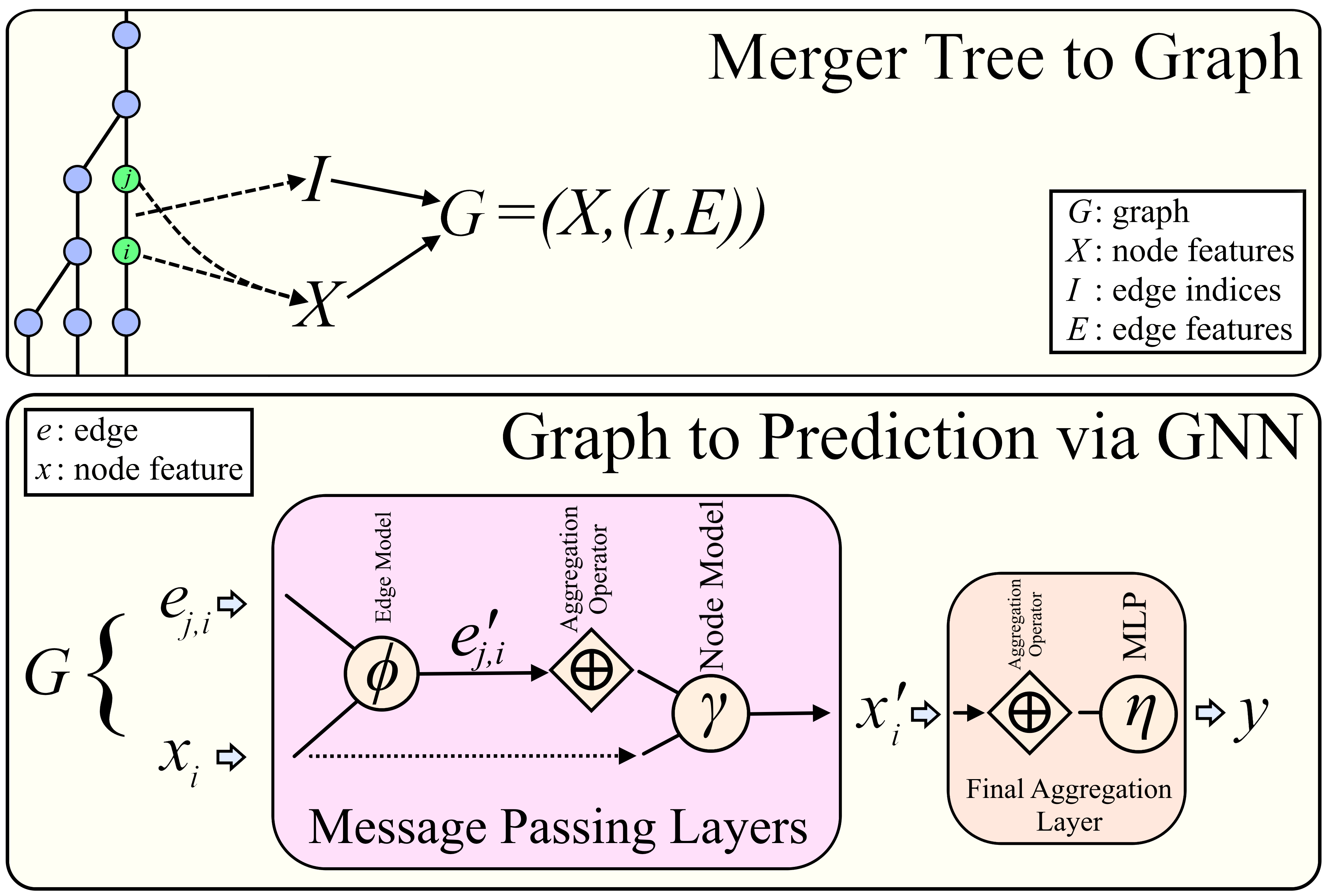}
    \caption{A schematic diagram of the tree-to-graph conversion and the GNN architecture. Node features and edges (consisting of edge indices and meaningless edge features) are encoded into a graph, which then pass through several message passing layers and a final aggregation layer, producing an output prediction.}
    \label{fig:diagram}
\end{figure}

\subsubsection{Architecture} \label{sec:architecture}

We use a GNN model similar to {\sc CosmoGraphNet}\footnote{https://github.com/PabloVD/CosmoGraphNet} \citep{2022zndo...6485804V} used in \cite{Villanueva-Domingo_2022}, which follows the Graph Network framework from \cite{47094}. \cite{Villanueva-Domingo_2022} use their GNN to both predict the power spectrum of a set of galaxies and infer the value of cosmological parameters. Thus, we choose to adopt the same GNN architecture with slight adaptations to facilitate compatibility of our merger tree graphs. We now briefly explain the model architecture as it applies to our analysis. For a more in-depth description of the architecture, we direct the reader to \cite{Villanueva-Domingo_2022}, Section~2.3.

The GNN consists of an encoder graph block, hidden message-passing layers, and a final aggregation layer. The encoder graph block and message-passing layers use a neighborhood aggregation scheme \citep[see][Section~2 for more details]{fey2019fastgraphrepresentationlearning}, incorporating edge and node models. Since we do not have edge features, the edge model exists only to satisfy the input requirements for the node model. In these layers, an edge model takes in the edge and node features and outputs new edge features, and a node model takes in the node and updated edge features and outputs new node features. Unlike \cite{Villanueva-Domingo_2022}, we do not input any global features into either model, as we are concerned solely with the structural properties of the merger tree and the data features of the individual subhalos, neither of which are described with global properties. Thus, the node model produces updated node features $x_i'$, such that:
\begin{equation} \label{eq:node model}
    x_i' = \gamma([x_i, ~\underset{j\in N_i}{\bigoplus} ~\phi(x_i,~x_j,~e_{j,i})])
\end{equation}
where $x_i, x_j$ are two linked nodes, $\phi$ and $\gamma$ are multilayer perceptrons (MLPs) which include a ReLU activation layer, $\phi(x_i,~x_j,~e_{j,i})$ represents the updated edge features, $N_i$ is the neighborhood of node $i$ (all the nodes with an edge connected to $i$), and $\bigoplus$ is a permutational invariant operator that aggregates all the information from neighboring nodes. Some examples of $\bigoplus$ are sum, mean, and maximum. 
We incorporate a multi-pooling layer in the node model to use all three possible aggregation operators before applying the MLP, which \cite{Villanueva-Domingo_2022} found produced slightly more accurate results. 

The message passing layers also include a residual layer which adds the input of the layer to its output; a technique that has been shown to reinforce the loss function against chaotic behavior \citep{NEURIPS2018_a41b3bb3}. The final aggregation layer uses the same multi-pooling layer with its aggregation operator as the node model, an MLP, and a final linear layer, which outputs an array of predictions $y$ for the parameter, such that:

\begin{equation} \label{eq:output}
    y=\eta ([\underset{i\in G}{\bigoplus}~ x'_i])
\end{equation}
where $\eta$ is an MLP, and $\bigoplus$ is defined as before. We show a diagram outlining this architecture as well as the graph construction in Figure~\ref{fig:diagram}.

\subsubsection{Node features} \label{sec:node features}
Of the available data features associated with each subhalo, we create combinations of five node features to attach to our graphs:

\begin{enumerate}
    \item \textbf{S}. The number of the snapshot in which the subhalo is found.
    \item $\mathbf{M_{DM}}$. Total mass of all DM particles that are bound to the subhalo.
    \item $\mathbf{M_{*}}$. Total mass of all stellar components that are bound to the subhalo.
    \item $\mathbf{M_{G}}$. Total mass of all gas components that are bound to the subhalo.
    \item \textbf{SFR}. The sum of the individual star formation rates of all gas cells in the subhalo.
\end{enumerate}

We use these node features in specific to summarize both the baryonic ($\mathbf{M_{G}}$, $\mathbf{M_{*}}$, \textbf{SFR}) and DM ($\mathbf{M_{DM}}$) attributes of subhalos. \textbf{S} is required in all combinations of node features to maintain the temporal information of the tree structure. 
Although including data from more {\sc Subfind} fields might produce more accurate inferences, we choose to maintain the focus of our work on testing the feasibility of our methods rather than optimizing quantitative results. For in-depth identification of WDM deterministic node features with GNNs, we direct the reader to Costanza et al. In Preparation.

\subsubsection{Loss Function}

We use the same loss function used to perform likelihood-free inference of cosmological parameters by \cite{Villanueva-Domingo_2022}. Our network is trained to predict the posterior mean and standard deviation of a parameter, which is done using a specific loss function given by \cite{jeffrey2020solvinghighdimensionalparameterinference}:

\begin{multline} \label{eq:loss}
    L \;=\; \log \left( \sum_{i=1}^N \sum_{\alpha=1}^A 
        \bigl(\theta_{\alpha,i} - \mu_{\alpha,i}\bigr)^2 \right) \\
    + \log \left( \sum_{i=1}^N \sum_{\alpha=1}^A 
        \left[\bigl(\theta_{\alpha,i} - \mu_{\alpha,i}\bigr)^2 
        - \sigma_{\alpha,i}^2\right]^2 \right) ,
\end{multline}

where $N$ are batch elements, $A$ are components, $\theta_{a}$ is the parameter, $\mu_a$ is the posterior mean, and $\sigma_a$ is the standard deviation.

\subsubsection{Training} \label{sec:training}

We divide our merger tree dataset into 80\% training, 10\% validation, and 10\% testing, using a seeded shuffle. We use a seeded shuffle because it means each unique GNN has identical dataset division, thus making comparisons between networks fair.  
Our GNN has four hyperparameters that we tune using {\sc Optuna} \citep{10.1145/3292500.3330701}. 
Automated tuning is necessary given the sensitivity of GNN performance to, e.g., architecture depth and learning rates. 
{\sc Optuna} does this by suggesting hyperparameters within a range for each trial that minimize the validation loss, meaning that each trial has a different set of values for these hyperparameters. The hyperparameters and their ranges are: learning rate (\textit{lr)} $10^{-7}$ to $10^{-3}$, weight decay (\textit{wd)} $10^{-9}$ to $10^{-6}$, number of layers (\textit{nl)} 1 to 5, and hidden channels (\textit{hc)} $2^n,$ with $n$ from 6 to 9. We train each of our models for 50 trials of 1000 epochs of hyperparameter tuning for each unique combination of node features.

\subsubsection{Evaluation Metrics} \label{sec:metrics}

We use two metrics to evaluate the success of our model. The coefficient of determination is our primary metric, which is given by: 
\begin{equation} \label{eq:r2}
    R^2 = 1 - \frac{\sum_i^N (y_{\rm true,i}-y_{\rm pred,i})^2}{\sum_i^N (y_{\rm true,i}-\bar{y}_{\rm pred,i})^2},
\end{equation}
where values closer to 1 mean more accurate results. We also calculate the root mean squared error (RMSE) as a secondary metric, defined such that:
\begin{equation} \label{eq:rmse}
    {\rm RMSE} = \sqrt{\frac{1}{N}\sum_{i}^{N}(y_{\rm true,i}-y_{\rm pred,i})^{2}}~,
\end{equation}
with smaller values meaning more accurate results. In both equations, $N$ is the number of trees in the testing set, $y_{\rm true,i}$ is the true parameter value, $y_{\rm pred,i}$ is the predicted parameter value, and $\bar{y}_{\rm pred,i}$ is the average of the predicted parameter values. We include both $R^2$ and RMSE since they describe different metrics for the ``accuracy'' of the model (the correlation of variables and error, respectively). However, as both the $R^2$ and RMSE similarly and sufficiently evaluate the GNN's performance, we generally defer to showing just the $R^2$ in the text. For completeness, we show both metrics in figures and tables.

\section{Results} \label{sec:results}

In this section, we present our results from training four GNN models to infer the cosmological parameter $M_\text{WDM}$ and astrophysical parameters $A_{\text{SN1}}$, $A_{\text{SN2}}$, and $A_{\text{AGN}}$. We first train the models with just one node feature, and then with increasingly more node features. We focus on the results of the $M_\text{WDM}$ model, as it performs the best and gives us the most interesting results. For extended results of the other three models, see Appendix \ref{app:other param results}.

\subsection{Merger History as an Input to GNNs} 
\label{sec:allparamsnap}

We begin testing our four GNN models by training them with just one node feature: snapshot number \textbf{S}. 
As the halos in merger trees evolve with time, having time dependence in the node features is essential. Including snapshot number accomplishes this, though notably, since we compute summary statistics across nodes, the snapshot number represents temporal information that is not necessarily the assembly of a halo with time. 
By training a model with the snapshot number alone, we are also implicitly testing whether the number and/or frequency of mergers in a given halo's history carries information about the underlying DM and astrophysics of our simulations. We explicitly test these sensitivities later in Section~\ref{sec:structuralinfo}.

\begin{figure*}
    \centering
    \includegraphics[width=\linewidth,trim={1.5cm 2cm 1.5cm 2cm},clip]{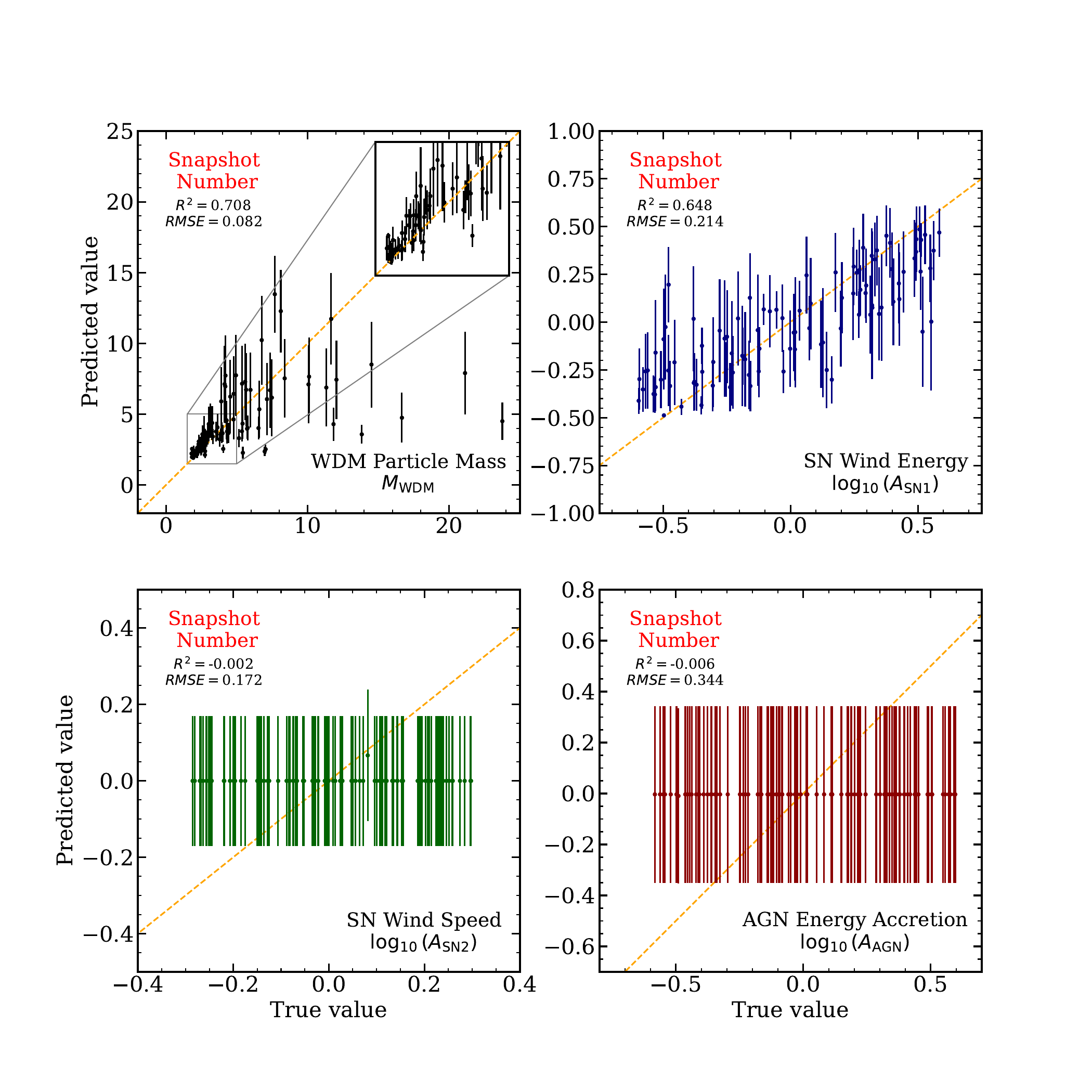}
    \caption{The plotted results of training the four GNN models on merger trees with one node feature attached: subhalo snapshot number. Each plot shows the inferred parameter values versus the true parameter values. The parameter inferred is labeled at the bottom right of each plot. The keys at the top left of each plot refer to the node feature that was used in training the model to produce the plot (see Section~\ref{sec:node features} for details). The dotted lines spanning the diagonals are where perfectly accurate predictions would fall. The inference of the WDM parameter is successful, particularly for lower masses, with a $R^2$ value of 0.708 and a RMSE of 0.082, as well as the SN1 parameter, with a $R^2$ value of 0.648 and a RMSE of 0.214. However, the $A_\text{SN2}$ and $A_\text{AGN}$ models both fail to make accurate predictions, as they only predict the mean value.}
    \label{fig:allparamsnap}
\end{figure*}

We train each model for 50 trials of 1000 epochs of hyperparameter tuning following the process in Section~\ref{sec:training}. 
Figure~\ref{fig:allparamsnap} shows the inferred parameter values compared against the true parameter values. 
Each plot shows the results from a different model (labeled in the bottom right of each panel).
The dotted lines show the one-to-one relation: models that are closer to this relation are more accurately predicting the true values. 

The top left panel of Figure~\ref{fig:allparamsnap} shows the results of the WDM GNN. We find that the GNN appears to successfully infer WDM particle mass for masses $\lesssim 4~{\rm keV}$. The box in the top right of the plot shows a zoom-in of this low mass regime. Although there is some variance, the general trend in this regime roughly follows the diagonal, demonstrating its success.
The success of the GNN in predicting WDM particle mass at $\lesssim 4~{\rm keV}$ is remarkable given that the input features of the graph is only the snapshot number.
This suggests that the number and/or frequency of merging events that a halo undergoes in its assembly is sensitive to the underlying DM physics.
Indeed, the sensitivity of the merger rates to DM physics makes qualitative sense, as many of the progenitors of the $z=0$ halos are low mass systems (see, e.g., Figure~\ref{fig:tree}).
The growth of these low mass systems is being preferentially suppressed by the initial matter power spectrum variations.
After $\sim 8~{\rm keV}$, however, the model consistently predicts WDM masses below the actual values.
We suspect that the model cannot do well in the $\gtrsim8~{\rm keV}$ range due to constraints from the simulation resolution, which we discuss further in Section~\ref{sec:massthresh}.

The top right panel of Figure~\ref{fig:allparamsnap} shows the results from inferring $A_\text{SN1}$. We find that the GNN infers the supernova wind energy parameter with some accuracy, making predictions that generally follow a diagonal trend, though with a relatively large degree of variance. Unlike the WDM inference, the model does about the same across the entire parameter range. The bottom left and bottom right panels show the results from inferring $A_\text{SN2}$ and $A_\text{AGN}$, respectively. Both plots show the same behavior: the models predict the mean value repeatedly (with the exception of a single simulation in the $A_\text{SN2}$ panel). This demonstrates a complete failure of these models to infer their parameters from the information we have trained them with. 

The lack of success in our model predicting the both the $A_{\rm SN2}$ and $A_{\rm AGN}$ parameters is perhaps unsurprising.
In low mass halos and at early times, the impact of variations in the supernova winds ($A_{\rm SN2}$) is curtailed by the minimum wind speed of the TNG model (see Equation~\ref{eq:asn2}).
The lack of distinction in the wind speeds at high redshift and in low mass halos means that many of the DM halos in our trees should be entirely unimpacted by variations in $A_{\rm SN2}$.
Similarly, black holes are not seeded in DM halos below a halo mass of $\sim7\times10^7 M_\odot$ \citep{10.1093/mnras/sty1733}.
AGN feedback is therefore not present in a large number of halos in the trees.
Then, once AGN feedback is present in the halos, the $A_{\rm AGN}$ variations play a self-regulatory effect on the total feedback budget from the black holes (see Appendix A of Garcia et al. In Preparation), making $A_{\rm AGN}$'s impact on the DM halos weak.
Therefore, we do not expect significant variations to the DM merger events by modifying either $A_{\rm SN2}$ or $A_{\rm AGN}$, consistent with our results.
However, as we explore in the next Section, $A_{\rm SN2}$ and $A_{\rm AGN}$ do become more important for the {\it baryonic} properties of the galaxies.

\begin{figure*}
    \centering
    \includegraphics[width=\linewidth,trim={1.5cm 2cm 1.5cm 2cm},clip]{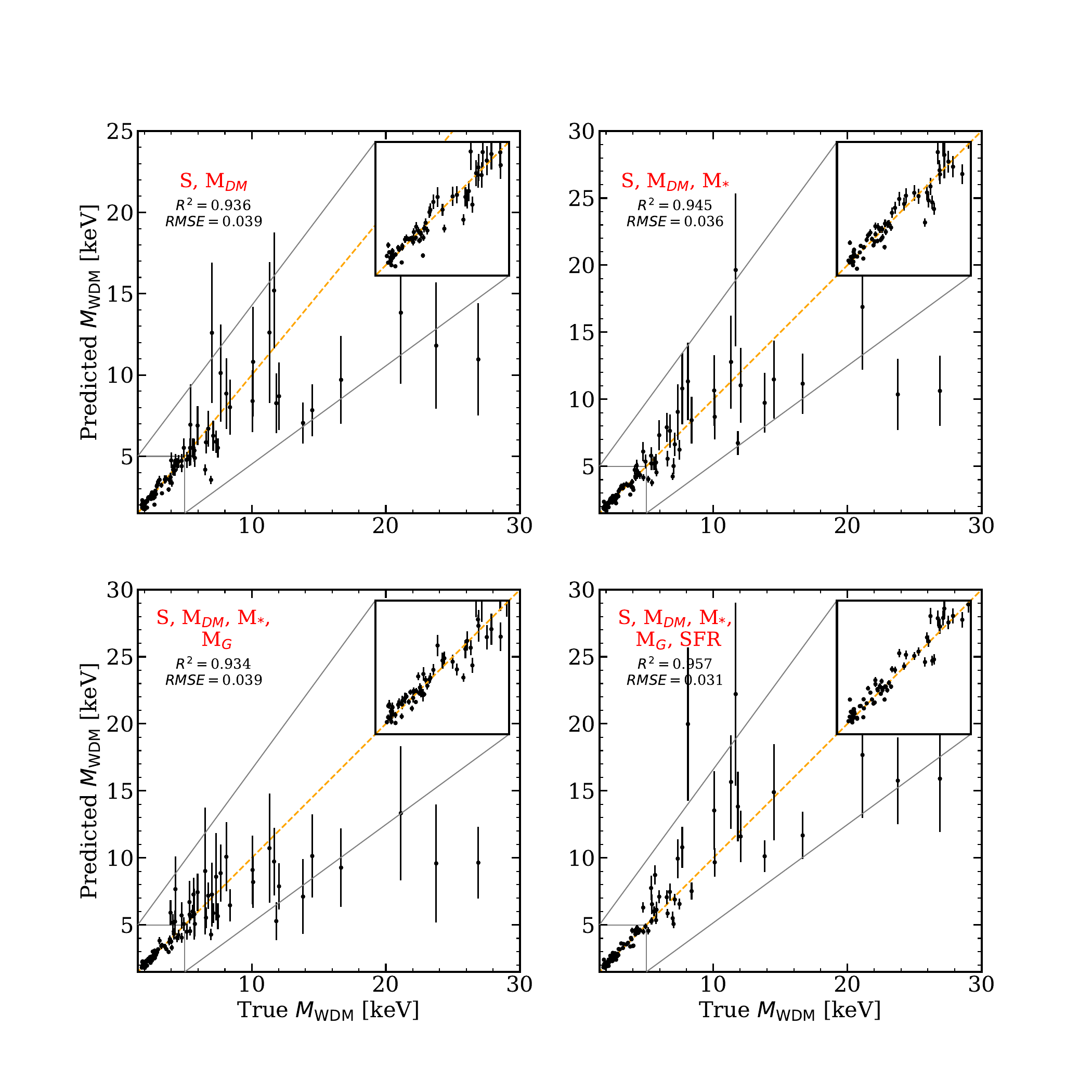}
    \caption{The plotted results of training the WDM GNN on four different combinations of node features from the merger trees. Each plot shows the inferred WDM particle mass versus the true WDM particle mass. The keys at the top left of each plot refer to the node features that were used in training the model to produce the plot (see Section~\ref{sec:node features} for details). The dotted line spanning the diagonals are where perfectly accurate predictions would fall. The model successfully infers the WDM particle mass for all of the node combinations, performing roughly the same for each combination of node features. The best performance, in the bottom right, has a $R^2$ value of 0.957 and a RMSE of 0.031, demonstrating highly accurate predictions.}
    \label{fig:alltrial_edit}
\end{figure*}

We quantify the performance of the models by calculating the coefficients of determination (see Section~\ref{sec:training}). The results from the $M_{\rm WDM}$ model produce a $R^2$ value of 0.708, $A_\text{SN1}$ produces a $R^2$ value of 0.648, $A_\text{SN2}$ produces a $R^2$ value of $-0.002$, and $A_\text{AGN}$ produces a $R^2$ value of $-0.006$. These values support the findings above, showing that only the $M_{\rm WDM}$ and $A_\text{SN1}$ models successfully learn to infer their parameters with some accuracy.

\subsection{Addition of Subhalo Features to GNNs} \label{sec:inferring wdm}

The previous subsection indicates that the structure of the merger tree histories themselves (i.e. the trees labeled only with snapshot number) contains some information about the WDM particle mass as well as the supernova wind energy ($A_{\rm SN1}$), but limited-to-no discriminatory information about the other astrophysical parameters.  
In this subsection, 
we investigate GNNs trained with additional node features which reflect galaxy properties that are more directly influenced by the assumed astrophysics. 
We first examine the benefit of the addition of physical properties of the galaxy and host halo in the GNN's ability to predict the WDM particle mass.
Then, using the same procedure, we present the best-performing models for the astrophysics parameter variations. 

We train our WDM GNN model with four progressively richer combinations of the node features listed in Section~\ref{sec:node features}, as follows: (\textbf{S}, $\mathbf{M_{DM}}$), (\textbf{S}, $\mathbf{M_{DM}}$, $\mathbf{M_{*}}$), (\textbf{S}, $\mathbf{M_{DM}}$, $\mathbf{M_{*}}$, $\mathbf{M_{G}}$), (\textbf{S}, $\mathbf{M_{DM}}$, $\mathbf{M_{*}}$, $\mathbf{M_{G}}$, \textbf{SFR}). We choose the combinations such that DM properties ($\mathbf{M_{DM}}$) are added first, and then baryonic properties ($\mathbf{M_{*}}$, $\mathbf{M_{G}}$, \textbf{SFR}). For each node combination, our network is trained for 50 trials of 1000 epochs of hyperparameter tuning. 

Figure~\ref{fig:alltrial_edit} shows the performance of the GNN trained on merger trees with the above node feature combinations.
Each panel of Figure~\ref{fig:alltrial_edit} shows the predicted WDM particle mass values against the actual WDM parameter values. 
Across all configurations, the models perform best in the low WDM particle mass regime ($\lesssim4~\mathrm{keV}$), where predictions closely track the true values with relatively low scatter. At larger WDM masses, the performance degrades for all networks with inferred values systematically undershooting the true values (consistent with the results of the previous section). At large WDM masses, all networks tend to predict WDM masses clustering around $\sim10$-$15~\mathrm{keV}$.  
This saturation effect is consistent with the resolution-driven limitations seen in earlier DREAMS analysis (expanded on in Section \ref{sec:massthresh}) and likely reflects the difficulty of distinguishing high-mass WDM models apart from one another where the suppressed portion of the matter power spectrum is not resolved \citep{10.1093/mnras/stad3260}. 

Increasing the richness of the node features improves the network performance in several ways.  
The first panel demonstrates that adding $\mathbf{M_{DM}}$ dramatically improves the accuracy of the model (when compared with Figure~\ref{fig:allparamsnap}), suggesting that including DM subhalo mass helps the model capture information about structure-suppression that is not encapsulated through the overall merger tree structure itself.
In the second panel, we see that adding $\mathbf{M_{*}}$ to the (\textbf{S}, $\mathbf{M_{DM}}$) configuration seems to produce a plot qualitatively similar to the previous, indicating it has little effect on the accuracy of the inferred values.
In the third panel, we incorporate $\mathbf{M_{G}}$, which reduces the prevalence of over-predictions (i.e., above the diagonal) for masses $\gtrsim10~\mathrm{keV}$. 
Finally, in the fourth panel, we add \textbf{SFR}, which generally increases the predicted values for masses $\gtrsim9~\mathrm{keV}$, which brings some predictions closer to the diagonal but also causes some large over-predictions around $\sim9~\mathrm{keV}$. Overall, it does not significantly alter the qualitative results.
These trends generally demonstrate that augmenting the node feature set with baryonic and DM halo properties progressively improves the GNN's ability to identify the employed WDM model, though the fundamental limitations associated with predicting WDM properties for the largest WDM mass scales remain.

We quantify the success of our network across each node combination by calculating the coefficients of determination (see Equation \ref{eq:r2}). We obtain coefficients of determination of 0.936 from (\textbf{S}, $\mathbf{M_{DM}}$), 0.945 from (\textbf{S}, $\mathbf{M_{DM}}$, $\mathbf{M_{*}}$), 0.934 from (\textbf{S}, $\mathbf{M_{DM}}$, $\mathbf{M_{*}}$, $\mathbf{M_{G}}$), and 0.957 from (\textbf{S}, $\mathbf{M_{DM}}$, $\mathbf{M_{*}}$, $\mathbf{M_{G}}$, \textbf{SFR}).
These results support our interpretation that the GNN is very successful across all four node combinations, though they also highlight a quantitative decrease in performance with the addition of $\mathbf{M_{G}}$. As this decrease is very slight, we attribute it to the GNN's variance from testing to testing, and not necessarily as an indication that this specific feature has detrimental effects on the GNN. Overall, both the qualitative and quantitative metrics indicate that the GNN successfully learns to infer WDM particle mass with a high degree of accuracy.

\begin{figure*}
    \centering
    \includegraphics[width=\linewidth,trim={3.5cm 0 3.5cm 0},clip]{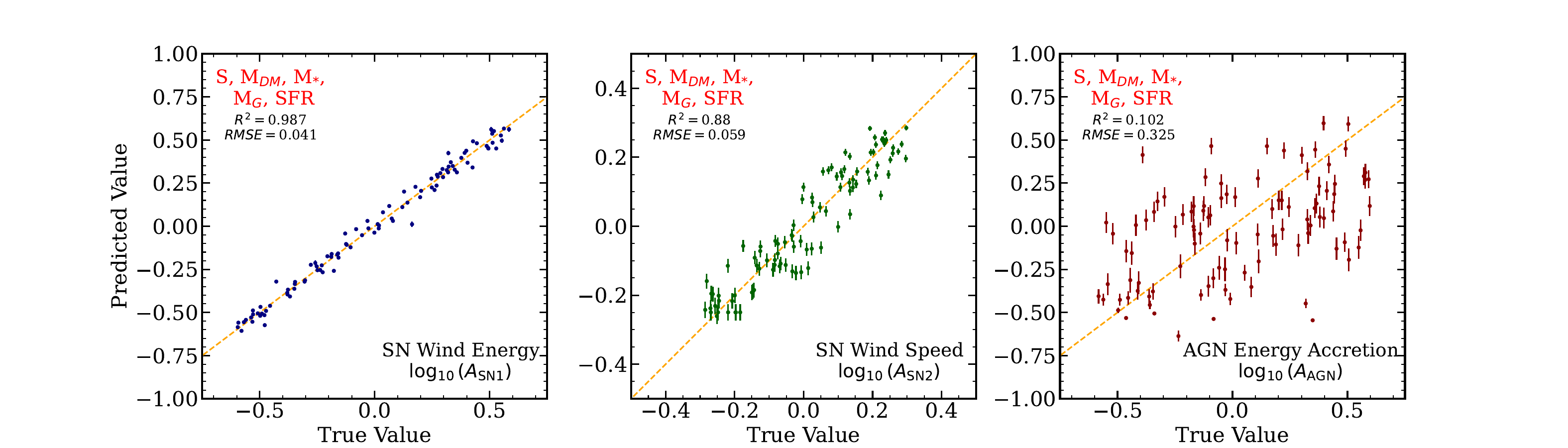}
    \caption{The plotted results of the best-performing node feature combinations from the $A_\text{SN1}$, $A_\text{SN2}$, and $A_\text{AGN}$ GNN models. Each plot shows the inferred parameter value versus the true parameter value in logspace. The dotted line spanning the diagonal is where perfectly accurate predictions would fall. In the top left of each plot, we show the node features used, and in the bottom right of each plot we show the parameter inferred by the model. The $A_\text{SN1}$ and $A_\text{SN2}$ models are successful ($R^2$ of 0.987 and $R^2$ of 0.880, respectively), and the $A_\text{AGN}$ model is unsuccessful ($R^2$ of 0.102).} 
    \label{fig:paramtrialfig}
\end{figure*}

Following the same procedure described above, we trained additional GNN models to infer the three astrophysical feedback parameters: 
$A_{\mathrm{SN1}}$, 
$A_{\mathrm{SN2}}$, and 
$A_{\mathrm{AGN}}$ (see Section~\ref{sec:parameters} for definitions).
For brevity, we present only the best-performing node feature combinations here, but the results of each node feature combination can be found in Appendix \ref{app:other param results}.
Each model underwent 50 trials of hyperparameter tuning over 1,000 epochs, using the same four node-feature combinations as in our WDM experiments: (\textbf{S}, $\mathbf{M_{DM}}$), (\textbf{S}, $\mathbf{M_{DM}}$, $\mathbf{M_{*}}$), (\textbf{S}, $\mathbf{M_{DM}}$, $\mathbf{M_{*}}$, $\mathbf{M_{G}}$), (\textbf{S}, $\mathbf{M_{DM}}$, $\mathbf{M_{*}}$, $\mathbf{M_{G}}$, \textbf{SFR}). The best-performing results for each parameter are presented in Figure~\ref{fig:paramtrialfig}. We also show the hyperparameters used in each result in Table~\ref{tab:hyperparameters}.

The networks show a mixed ability to recover these feedback parameters. For $A_{\mathrm{SN1}}$, the GNN achieves strong predictive performance ($R^{2} = 0.987$) for (\textbf{S}, $\mathbf{M_{G}}$, $\mathbf{M_{*}}$, $\mathbf{M_{G}}$, \textbf{SFR}), indicating that galaxy properties such as stellar mass and star formation rate are sensitive to supernova wind energy injection. 
The $A_{\mathrm{SN2}}$ model performs slightly worse ($R^{2} = 0.880$) for (\textbf{S}, $\mathbf{M_{G}}$, $\mathbf{M_{*}}$, $\mathbf{M_{DM}}$, \textbf{SFR}), but still shows a clear correlation between predicted and true values. 
This suggests that signatures of the supernova wind velocity scaling are present, albeit with slightly increased variance compared to $A_{\mathrm{SN1}}$. 
In contrast, the model for $A_{\mathrm{AGN}}$ fails to learn meaningful relationships ($R^{2} = 0.102$) with any node combination, with predictions not following any trend. 
This lack of predictive power is consistent with expectations: our Milky Way–mass zoom-in suite predominantly samples halos below the stellar-mass scale where AGN feedback becomes important \citep{10.1093/mnras/sty1733}, limiting the amount of $A_{\mathrm{AGN}}$ information encoded in the subhalo merger histories.

\begin{table*}
    \centering
    \begin{tabular}{llllllll}
        \multicolumn{7}{c}{Best Performances} \\
        \hline
        Parameter&$R^2$&RMSE&\textit{lr}&\textit{wd}&\textit{nl}&\textit{hc}\\
        \hline
         $M_\text{WDM}$&0.957&0.031 &$1.889\cdot10^{-5}$ &$5.139\cdot10^{-7}$ &4 &128 \\
         $A_\text{SN1}$&0.987&0.041 &$7.862\cdot10^{-6}$ &$4.352\cdot10^{-9}$ &5 &64 \\
         $A_\text{SN2}$&0.880&0.059 &$1.063\cdot10^{-7}$ &$2.826\cdot10^{-9}$ &4 &256 \\
         $A_\text{AGN}$&0.102 &0.325 &$1.778\cdot10^{-74}$ &$1.196\cdot10^{-7}$ &5 &128  
    \end{tabular}
    \caption{Metrics of performance and hyperparameters from the best trial of the best-performing node combination for each parameter. The hyperparameters are learning rate \textit{lr}, weight decay \textit{wd}, number of layers \textit{nl}, and hidden channels \textit{hc}.}
    \label{tab:hyperparameters}
\end{table*}

\section{Discussion} \label{sec:discussion}

In this section, we conduct additional tests of merger tree structure and discuss the trends in our results from inferring the four parameters.

\subsection{Merger Tree Structure without Temporal Information} \label{sec:structuralinfo}

The results of our WDM GNN trained on just one node feature (see Section~\ref{sec:allparamsnap}) demonstrate that the network can infer WDM particle mass with snapshot number as the only attached information. The snapshot number represents the merger tree's time dependence, essentially assigning a time to each node feature. 
By including no other node features, the network is being trained on the structure of the merger tree histories alone which includes, for example, the total number of subhalos/mergers that contributed to the formation of the $z=0$ halo. 
The success of the network with the absence of additional node features suggests that there is information contained in the structure of a subhalo's merger tree history regarding the WDM particle mass within the subhalo.

\begin{figure}
    \centering
    \includegraphics[width=\linewidth]{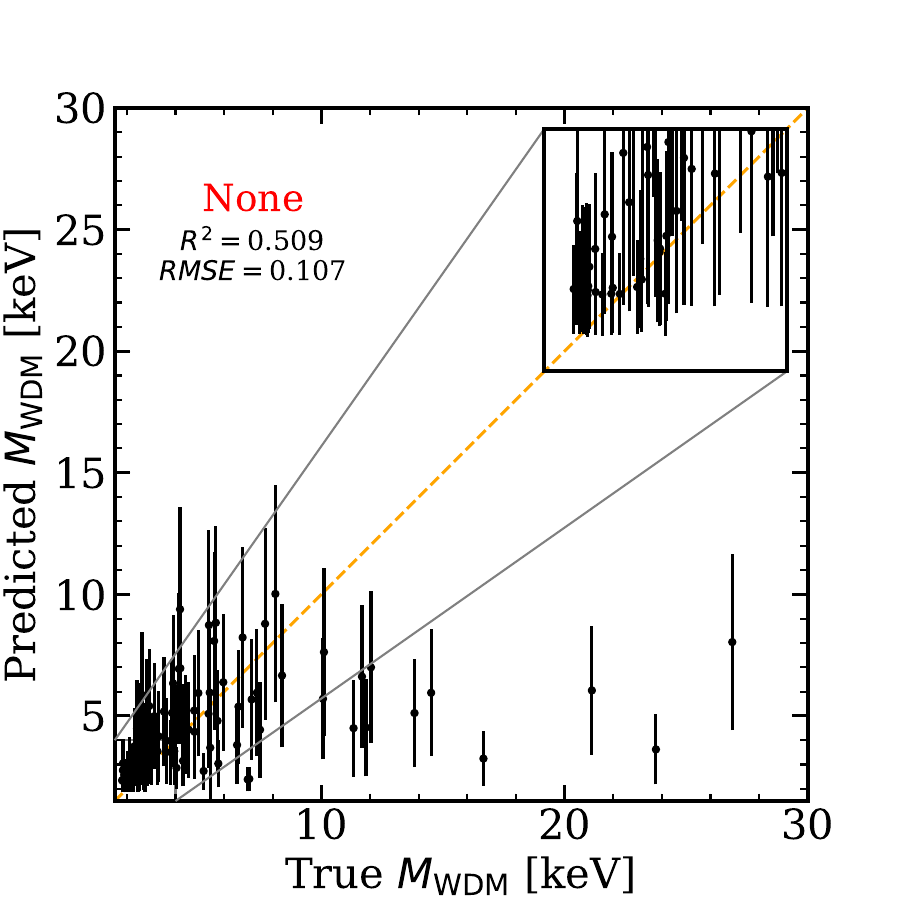}
    \caption{The plotted results of training the WDM GNN without the addition of any node features. The plot shows the inferred WDM particle mass versus the true WDM particle mass. The dotted line across the diagonal represents perfectly accurate predictions. The model makes inferences with a $R^2$ value of 0.509 and a RMSE of 0.107, which demonstrates a small amount of success.}
    \label{fig:nonefig}
\end{figure}

We further investigate the merger tree structural information by training our WDM GNN on merger trees without time dependence. 
We do this by creating graph objects with only a placeholder node feature $x = [1, 1, \ldots ]$ attached to satisfy our GNN requirements. 
Without the addition of snapshot number as a node feature, we have stripped away the temporal element of the trees, thus training our GNN on the number of nodes in the graph and the associated edge connections. 
We train the model for 50 trials of 1000 epochs of hyperparameter tuning. 
The results are shown in Figure~\ref{fig:nonefig}.
We find that the model is able to infer the mass of the WDM particle, though with significantly less accuracy than previous results. It is particularly successful at low masses, but, unlike the results in Figure~\ref{fig:allparamsnap}, there is also high uncertainty in this regime.
Following the previously seen trend of the $M_{\rm WDM}$ GNNs, we also see it undershoot most of the predictions after $\sim7~{\rm keV}$. 

We calculate a $R^2$ value of 0.509 and a RMSE of 0.107. This is the least successful our WDM network has performed. Still, since we do not see a total failure of the model (like shown in the bottom panels of Figure~\ref{fig:allparamsnap}), this result signifies that there is some amount of meaningful information regarding the employed DM model of our simulations contained in the structure of the merger events, albeit not as much as with the addition of time dependence.

\subsection{WDM Mass Threshold} \label{sec:massthresh}

Our $M_{\rm WDM}$ model was generally accurate in its predictions at low masses, but at greater masses, the accuracy decreased and the uncertainties increased. As shown in the top two panels and bottom left panel of Figure~\ref{fig:alltrial_edit}, after a mass threshold of $\sim8~{\rm keV}$, the network begins to undershoot the true parameter values consistently, oftentimes making predictions clustered around $\sim10~{\rm keV}$. This result is qualitatively similar to \cite{10.1093/mnras/stad3260}, who train a CNN to infer WDM mass on full box N-body simulations from images with limited success in similar mass ranges (see their Figures 4-6). They attribute this to the resolutions of their simulations, which restricts the maximum WDM mass that is distinguishable. We are similarly constrained by simulation resolution, and so after a certain mass our GNN cannot easily distinguish between WDM models. In our case, this mass is $\sim8~{\rm keV}$.

\subsection{AGN Model Performance} \label{sec:agnperformance}

The results from the $A_\text{AGN}$ model shown in Figure~\ref{fig:paramtrialfig} demonstrate that the model cannot infer the AGN feedback parameter, regardless of the node features it learns from. These results match \cite{Lin_2024} (see their Figure~2), who train a conditional normalizing flow on 14 node features from individual DREAMS galaxies at $z=5$ to infer the same parameters as us. Our $R^2$ values in Section~\ref{sec:inferring wdm} are also quantitatively similar to the results in their Section~3.1. 

We speculate that the inability of our model to infer $A_\text{AGN}$ is due to the MW zoom-in simulation suite we have selected for our dataset. The AGN feedback parameter consists of a high-accretion state and a low-accretion state, which are active for different mass ranges. For stellar masses $\lesssim10^{10.5}~ {\rm M_\odot}$, the low-accretion state is not active, and therefore does not influence the MW zoom-in suite, as mostly all the subhalos in this suite fall below this mass \citep{10.1093/mnras/sty1733}. Since our dataset is comprised of data features from these subhalos, there is likely limited information within it regarding $A_\text{AGN}$ that our GNN can learn from.

\subsection{Pruned Trees} \label{sec:prune}

Results from \cite{jespersen2022mangrove} indicate that reducing the noise of our data could improve the ability of our models to infer parameters. In this section, we experiment with pruning merger trees, that is, removing certain nodes and branches from the trees to make the data less noisy. We try three methods of removing nodes and branches based on three different criteria. The first method is removing all nodes with fewer than 100 DM particles. The second is removing all nodes with less than 100 DM particles that are also not significant in the tree structure (the nodes that are the first and last node in each branch are significant, as these are the nodes that define a branch of the tree). The third is removing all branches that begin with a node with less than 100 DM particles. Between these criteria, the third method produces the simplest and least noisy trees, so we choose to test this method.

\begin{figure}
    \centering
    \includegraphics[width=\linewidth]{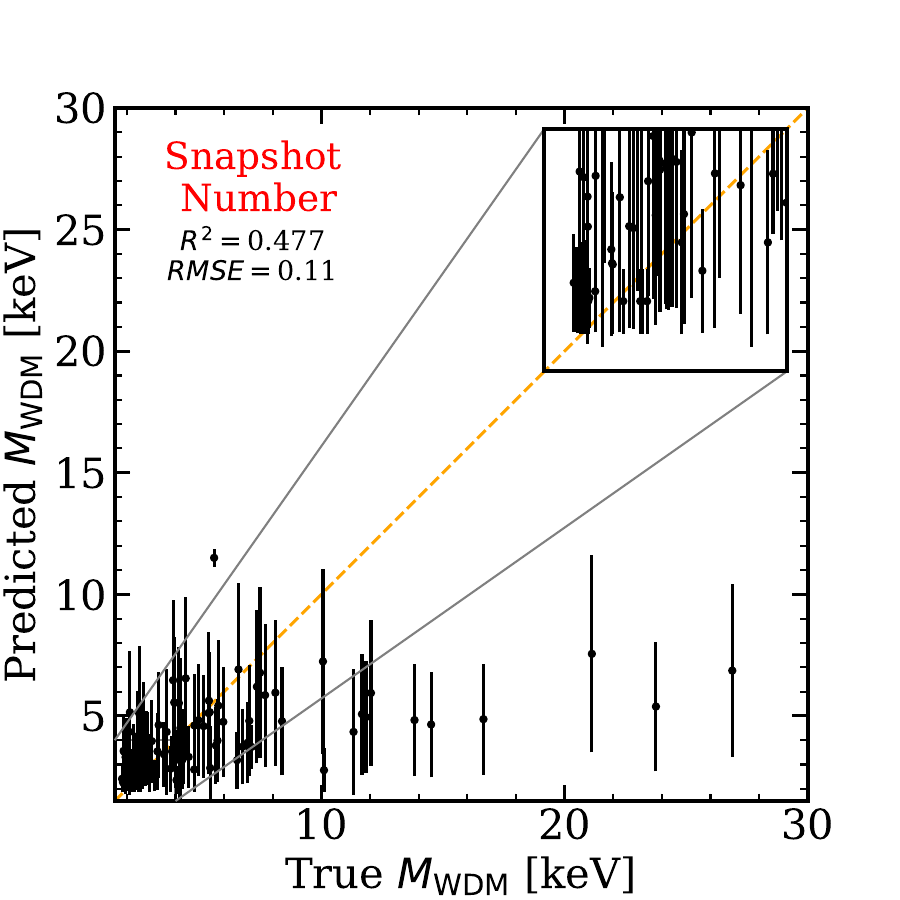}
    \caption{The plotted results of training the WDM GNN with pruned merger trees and snapshot number as a node feature. The plot shows the inferred WDM particle mass versus the true WDM particle mass. The dotted line along the diagonal is where perfectly accurate predictions would fall. The model is somewhat successful, with a $R^2$ value of 0.477 and a RMSE of 0.110, though less so than its un-pruned counterpart (shown in Figure~\ref{fig:allparamsnap}, top left).}
    \label{fig:prunefig}
\end{figure}

We train our WDM GNN with snapshot number as the only node feature from our merger trees pruned using the third method, for 50 trials of 1000 epochs of hyperparameter tuning. We show the results in Figure~\ref{fig:prunefig}. We find that the GNN is able to make accurate predictions, though less accurate than training it on the same node features with our original trees (shown in Figure \ref{fig:alltrial_edit}). We see that the model overshoots the true values for many points, until $\sim8~{\rm keV}$, after which all of the points are below the true values. We calculate a $R^2$ value of 0.454 a RMSE of 0.117, supporting our above findings. The $R^2$ value is less than the $R^2$ value from the same node combination without pruning the trees, which indicates that our WDM GNN model benefits from a larger selection of subhalos in our dataset.

\subsection{Sensitivity Tests} \label{sec:senstest}

\begin{figure*}
    \centering
    \includegraphics[width=\linewidth,trim={1cm 0 1cm 0},clip]{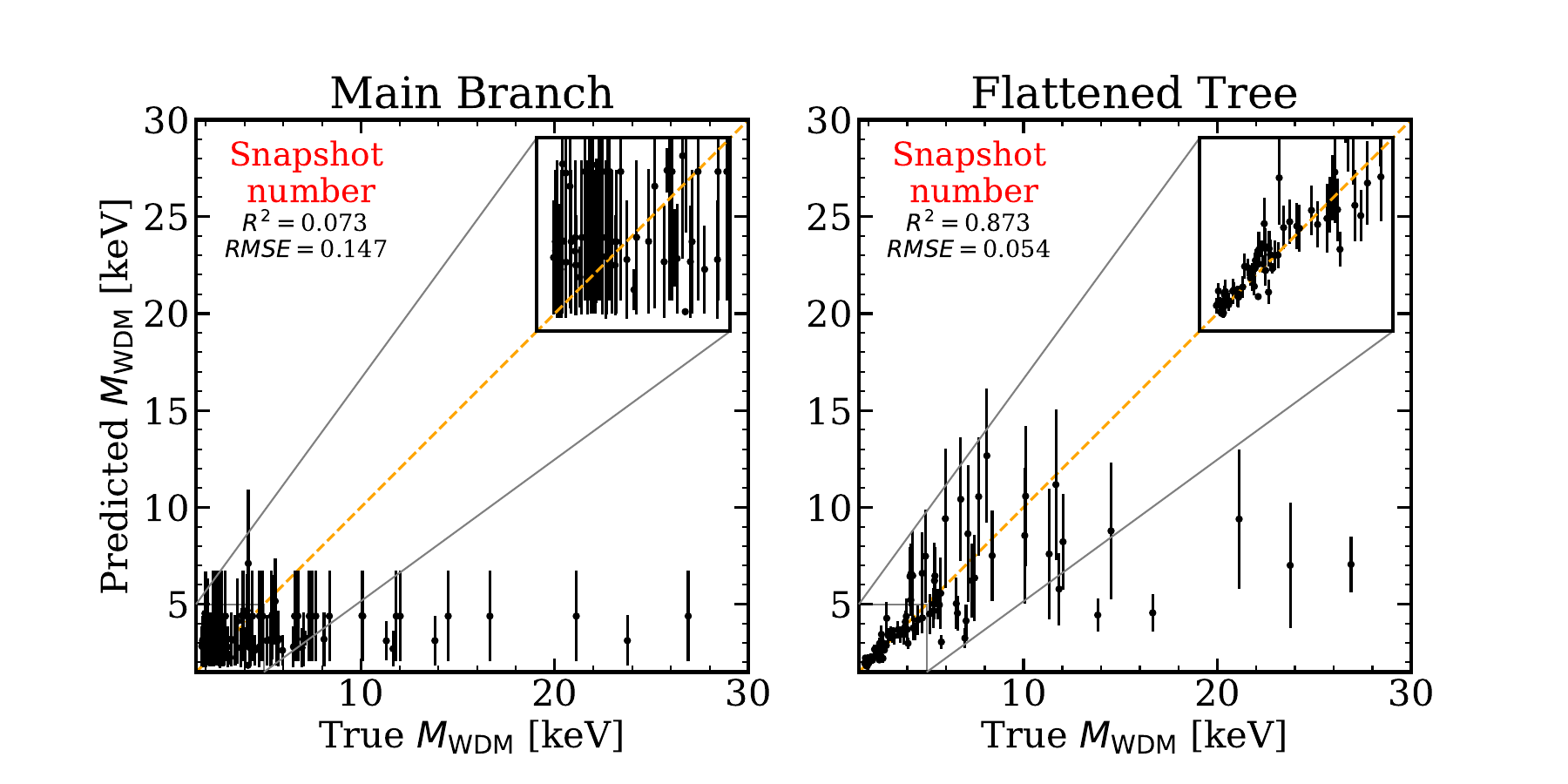}
    \caption{The plotted results of training the WDM GNN with just the main branch (left) and with all the subhalos but only using a single sequence of edges (right). In the first instance, the model fails to infer the WDM particle mass, but in the second, the model does learn to infer accurately, with a $R^2$ value of 0.873 and a RMSE of 0.054.}
    \label{fig:senstests}
\end{figure*}

To investigate which specific components of the merger trees are sensitive to WDM, we conduct two tests. First, we train the $M_\text{WDM}$ GNN with just the most massive progenitors stemming from the first subhalo (called the main branch) of each merger tree, and second with all the nodes in a given tree organized into a single long sequence of edges (flattened trees). We construct flattened trees by trivially drawing edges between consecutive nodes in the data (i.e., the node at index $i$ is connected to the node at index $i+1$) instead of matching through hereditary merger links, which effectively eliminates the merger information contained within edges. These tests show if the number of nodes (or, analogously, the number of subhalos) in the tree is more important for WDM mass inference than the edges connecting them. We train the $M_\text{WDM}$ model for 50 trials of 1000 epochs of hyperparameter tuning with snapshot number as the only node feature for both of these tests, and show the results in Figure~\ref{fig:senstests}.

In the left panel, we show the results from training the model on just the main branch. There is a clear non-trend in the predicted values, which indicates that the model is unsuccessful. These results have a of $R^2=0.073$ and a RMSE of $0.147$, supporting this interpretation. 
In the right-hand panel, we show the results from training the model on all the nodes, but in a flattened tree structure instead of the normal merger tree. 
Compared with the results from the original tree with the same node feature (top left panel of Figure~\ref{fig:allparamsnap}), the flattened trees qualitatively seem about the same. There still is a trend following the diagonal, particularly at lower masses, and the predicted values drop off after a certain mass threshold. This test has a $R^2$ value of 0.873 and RMSE of $0.054$, and the original tree produced a $R^2$ of 0.708 and a RMSE of 0.082. The difference in the quantitative metrics between the flattened and original trees does not demonstrate any extreme difference between the two results, as highlighted in the plots. As in Section~\ref{sec:inferring wdm}, we attribute this difference to variance in the GNN from testing to testing. We repeat these tests for the astrophysics GNNs as well (results shown in Appendix~\ref{app:other param results}), and find that the $A_\text{SN1}$ GNN cannot learn from just the main branch, but can with the flattened tree, similar to the WDM GNN. As the $A_\text{SN2}$ and $A_\text{AGN}$ models already cannot learn from the full tree with the snapshot number (shown in Section~\ref{sec:allparamsnap}), they expectedly are unable to make accurate predictions with either the main branch or flattened tree tests.

These results provide some insight as to where the WDM sensitivities lie within the merger trees. Our tests indicate that the number of subhalos in a merger tree is very sensitive to WDM, whereas the graph structure we use is less so. In comparing Figure~\ref{fig:allparamsnap}~(top left panel)~and~Figure~\ref{fig:senstests} to Figure~\ref{fig:alltrial_edit}, we also see that the DM halo mass has significant predictive power for the WDM mass. Indeed, previous work has found similar results: the halo mass function and progenitor mass functions have been shown to be sensitive to the free-streaming scale of DM and therefore WDM particle mass \citep{benson2013dark}. Though we do not directly measure these mass functions, by including $\mathbf{M_{DM}}$ as a node feature, the merger tree histories naturally gain information regarding these mass functions. This can explain why $\mathbf{M_{DM}}$ has such predictive power. Across all of our tests, the two strongest sensitivities to WDM mass seem to be the number of subhalos and the DM halo mass.

\subsection{Feature Importance} \label{sec:feature importance}

We conduct further tests into the WDM sensitivity of each feature using the boosted decision tree based ML algorithm called eXtreme-Gradient Boosting ({\sc XGBoost}) \citep{Chen_2016}. {\sc XGBoost} generates an ensemble of decision trees, comprised of a base tree and subsequent trees formed from the residual errors after each iteration. The contributions from these subsequent trees are scaled by the learning rate, and are summed to produce the final prediction from the model. The model is trained to minimize the mean squared error between WDM particle mass predictions and the corresponding true WDM values. Using subsequent learners in this way is more accurate and robust than using a single decision tree for the same purpose.  

Our primary interest with this test is evaluating feature importance, which is a metric of how much a given feature improves the performance of the model. This, combined with the previous tests in Section~\ref{sec:senstest}, let us compare how two independent ML models highlight WDM sensitivities in our data, giving deeper insight into what the most important components of our merger tree dataset in inferring WDM are. 

We create our {\sc XGBoost} model using nine features. Five of these features are the features we used with our GNN (\textbf{S}, $\mathbf{M_{DM}}$, $\mathbf{M_{*}}$, $\mathbf{M_{G}}$, \textbf{SFR}). 
Feature six is, per the results of Section~\ref{sec:senstest}, the number of nodes $\mathbf{N_{nodes}}$. 
Features seven through nine are ratios of gas mass, DM halo mass, and stellar mass ($\mathbf{M_{G}}$/$\mathbf{M_{DM}}$, $\mathbf{M_{G}}$/$\mathbf{M_{S}}$, $\mathbf{M_{S}}$/$\mathbf{M_{DM}}$), which we anticipate may contain more robust information than the individual mass function cutoffs. For each merger tree, we take all the nodes within the tree and compute summary statistics for each feature (e.g., the median, the 84th percentile). This means the model is given nine summary statistics corresponding to each feature for every merger tree in our dataset. We show the results from using the 84th percentile as our summary statistic for each feature, since we find it to produce the most accurate WDM predictions compared with other summary statistics. We use the same merger tree dataset used for our GNN (described in Section~\ref{sec:simulations}), though only including subhalos with $z>7$ and $10^{7.5}\text{M}_\odot<\mathbf{M_{DM}}<10^{8.5} \text{M}_\odot$. We make these restrictions to focus on WDM models with the most pronounced free-streaming effects and with the least interference from later-time astrophysical feedback processes. This dataset is divided with an 80/20 training/testing split.

Our model is trained with six hyperparameters that are tuned using {\sc RandomizedSearchCV} from the {\sc sklearn} library \citep{DBLP:journals/corr/abs-1201-0490}. The hyperparameters and their search space ranges are: the number of trees (\textit{ne}) 100 to 600, the depth of each tree (\textit{md}) 3 to 10, the learning rate (\textit{lr}) 0.01 to 0.2, the row sampling per tree (\textit{ss}) 0.4 to 0.6, the feature sampling (\textit{fs}) 0.4 to 0.6, and the minimum instance weights (\textit{mw}) 1 to 9. We randomly sample over these ranges for 50 iterations, which finds the best hyperparameters: \textit{ne} = 289, \textit{md} = 6, \textit{lr} = 0.019, \textit{ss} = 0.636, \textit{fs} = 0.760, \textit{mw} = 4.297.

\begin{figure}
    \centering
    \includegraphics[width=\linewidth,trim={0 0 2cm 0},clip]{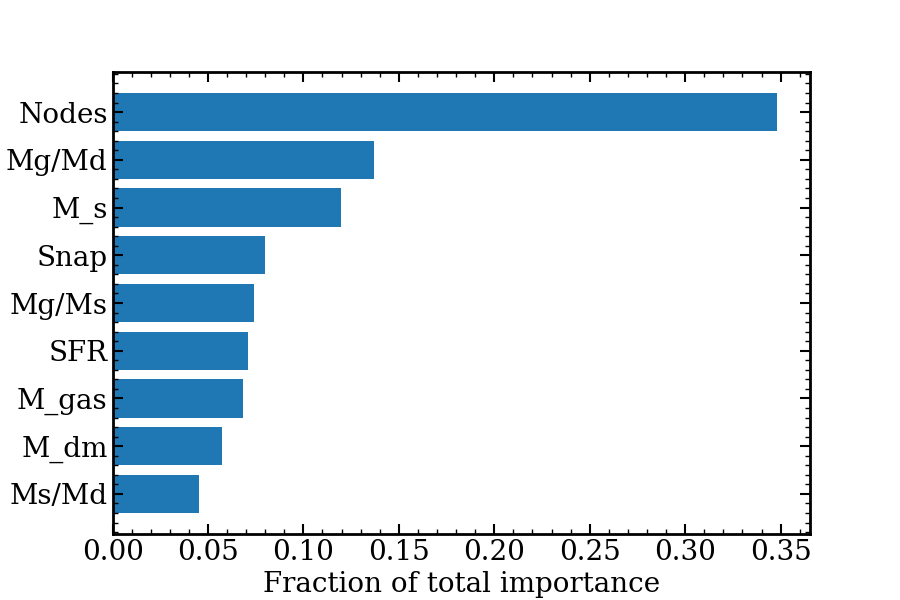}
    \caption{The fraction of total importance for the nine features used to train our {\sc XGBoost} model. The number of nodes is more than twice as important as the next most important feature, indicating its extreme sensitivity to WDM. Gas-to-DM mass ratio and stellar mass are the next two most important features. The remaining features are marginally important, contributing less than 10 percent each to the total.}
    \label{fig:xgb}
\end{figure}

With these hyperparameters, our model makes predictions of the WDM particle mass with a $R^2$ of 0.847 (successful, though less so than our GNN). We show the feature importance for the model in Figure~\ref{fig:xgb}. Clearly, the number of nodes is the most important feature, with an importance (0.348) of more than double the next feature. This supports our findings with the GNN tests (Section~\ref{sec:senstest})-- the total number of subhalos in a merger tree is a strong indicator of WDM. However, unlike with our GNN, the DM halo mass is not important to the XGBoost model, as it has the second lowest feature importance (0.057). It is possible this discrepancy between ML models arises from what specifically they are learning from: our GNN is trained on each individual subhalo DM mass, whereas the XGBoost model uses summary statistics of the total DM halo mass for a given merger tree. This indicates that the evolution of DM mass over time, rather than one summary statistic of DM mass, has more sensitivity to WDM. This is an expected result, as naturally, DM evolution encodes more information.

Nonetheless, the ratio of gas mass to DM mass has significantly more importance to the model, with a fractional importance of 0.137, making it the second most important of the seven features. The importance of this feature may indicate that subhalos in lower WDM universes are more gas rich relative to their DM mass.
Essentially, the gas-to-DM mass ratio seems to be a sensitive indicator of gas retention efficiency, which captures the behavior of subhalos that are most gas rich.
A possible interpretation of the importance of $M_{\rm g}/M_{\rm d}$ is that subhalos in colder DM (i.e., low $M_{\rm WDM}$) models are able to accrete and retain a larger fraction of cosmic baryons, making them more resilient to stripping effects from the UV background from reionization. For warmed DM (i.e., high $M_{\rm WDM}$) models, there is delayed and less efficient halo formation, making subhalos more strongly affected by gas losses and thus decreasing their gas-to-DM mass ratios.

\section{Conclusions} \label{sec:conclusions}

In this paper, we demonstrate that GNNs trained on galaxy merger trees can recover both astrophysical feedback parameters and the underlying DM model properties used within cosmological simulations. 
We trained several GNNs using an array of combinations of subhalo properties to assess which features were most informative for recovering each model parameter. 
Our conclusions are as follows:

\begin{itemize}
    \item We presented a modified version of the \textsc{CosmoGraphNet} architecture designed to take annotated galaxy merger trees as inputs and predict simulation parameters including specifically the adopted WDM particle mass and astrophysics feedback model parameters (Section~\ref{sec:methods}).
    
    \item We trained a series of GNNs using data from the DREAMS WDM MW zoom-in suite (Section~\ref{sec:inferring wdm})
    to infer WDM particle masses. Notably, we found that GNNs trained with only time information (snapshot number) were able to predict WDM masses (Section~\ref{sec:allparamsnap} and Figure~\ref{fig:allparamsnap}). While this could be attributed to WDM simply impacting the number of nodes within a tree, GNNs trained without time information failed to perform equivalently well (Section~\ref{sec:structuralinfo} and Figure~\ref{sec:structuralinfo}).  This indicates that not only the abundance but the temporal structure of the merger tree contains information about the adopted initial matter power spectrum. 

    \item We quantified how progressively enriching the node features (e.g., adding stellar mass, gas mass, DM mass, and star formation rate) improved the network’s ability to recover WDM particle masses, reducing prediction scatter and extending the range of WDM masses over which  accurate inference could be performed (Figure~\ref{fig:alltrial_edit}). 

    \item We identified a performance threshold for the WDM inference with GNNs reliably being able to recover the WDM masses in the range $\lesssim6$–$10~{\rm keV}$ but systematically underestimating higher-mass values. We attributed this limitation to the decreasingly well-resolved distinction between high-mass WDM models behavior at the resolution of our zoom-in simulations (Sections \ref{sec:results} and \ref{sec:massthresh}).

    \item We demonstrated that the same GNN inference can be used to predict the SN wind energy $A_\text{SN1}$ and the SN wind velocity $A_\text{SN2}$ astrophysical parameters, but cannot be used to predict the AGN feedback parameter $A_\text{AGN}$ parameter (Figure~\ref{fig:paramtrialfig}). We speculate that the performance of our $A_\text{AGN}$ model is due to the subhalo mass limitations in our MW zoom-in suite, which could potentially be resolved by using a dataset from a different simulation suite (Section~\ref{sec:agnperformance}).

    \item We investigated the sensitivity of WDM to different components of the merger trees. We determined that subhalo abundance in merger trees and the DM halo mass are dominant for WDM inference with our GNN (Sections~\ref{sec:prune},~\ref{sec:senstest}, Figures~\ref{fig:prunefig},~\ref{fig:senstests}), whereas the actual graph structure is less so. We directly investigated feature importance using an {\sc XGBoost} model, again showing subhalo abundance to be highly informative for the models (Section~\ref{sec:feature importance}, Figure~\ref{fig:xgb}). With this method, we also found significant sensitivity to the gas-to-DM halo mass ratio of subhalos, possibly indicating that colder DM models lead to increased subhalo resilience against stripping effects from the UV background during reionization (Section~\ref{sec:feature importance}, Figure~\ref{fig:xgb}).

\end{itemize}

This work offers a way to clarify how simulation assumptions shape galaxy properties and their formation histories. This framework can help identify which galaxy features are most sensitive to underlying cosmological and feedback parameters, guiding the design and interpretation of future simulation studies.

\section*{Acknowledgments}
The authors acknowledge Research Computing at The University of Virginia for providing computational resources and technical support that have contributed to the results reported within this publication. URL: \href{https://rc.virginia.edu}{https://rc.virginia.edu}.
PT and AG acknowledge support from NSF-AST 2346977. PT, AG, and AF the NSF-Simons AI Institute for Cosmic Origins (\url{http://cosmicai.org}) which is supported by the National Science Foundation under Cooperative Agreement 2421782 and the Simons Foundation award MPS-AI-00010515.

\section*{Data Access}
The code used to produce the results in this paper is publicly available on \href{https://github.com/ilemleisher/merger_tree_gnn}{GitHub}\footnote{\href{https://github.com/ilemleisher/merger_tree_gnn}{https://github.com/ilemleisher/merger\_tree\_gnn}}.

\bibliography{bibliography}{}
\bibliographystyle{aasjournal}

\appendix
\section{Astrophysical Parameters Results Extended} \label{app:other param results}

In this section, we show the results for all of the node feature combinations used with the astrophysical parameter models $A_\text{SN1}$, $A_\text{SN2}$, $A_\text{AGN}$ in Figure~\ref{tab:allparamresults}.

\tikzset{ 
table/.style={
  matrix of nodes,
  row sep=-\pgflinewidth,
  column sep=-\pgflinewidth,
  nodes={rectangle,draw=black,align=left},
  text depth=0.5ex,
  text height=1.75ex,
  nodes in empty cells
},
row 1/.style={nodes={fill=gray!10,align=left}},
column 1/.style={nodes={fill=gray!10,text width=1cm,align=right}}
}
\begin{figure}[h]
    \centering
    \begin{tikzpicture}

    \matrix (mat) [table,   
        column 1/.append style={minimum width=1.0cm},
        column 2/.append style={minimum width=2.3cm},
        column 3/.append style={minimum width=2.3cm},
        column 4/.append style={minimum width=3.2cm},
        column 5/.append style={minimum width=4.2cm},
        column 6/.append style={minimum width=2.3cm},
        column 7/.append style={minimum width=2.3cm}]
    {
    & \textbf{S}, $\mathbf{M_{G}}$ & \textbf{S}, $\mathbf{M_{G}}$, $\mathbf{M_{*}}$ & \textbf{S}, $\mathbf{M_{G}}$, $\mathbf{M_{*}}$, $\mathbf{M_{DM}}$ & \textbf{S}, $\mathbf{M_{G}}$, $\mathbf{M_{*}}$, $\mathbf{M_{DM}}$, \textbf{SFR}&MB&FT \\
    $A_\text{SN1}$ & 0.737, 0.185 & 0.966, 0.066 & 0.953, 0.078 & 0.987, 0.041 & 0.021, 0.356 & 0.703, 0.169 \\
    $A_\text{SN2}$ & $-0.003$, 0.172 & 0.541, 0.116 & 0.837, 0.069 & 0.880, 0.059 & $-0.003$, 0.172& $-0.013$, 0.173 \\
    $A_\text{AGN}$ & $-0.012$, 0.345 & $-0.013$, 0.345 &$-0.005$, 0.344  & $0.102$, 0.325 & $-0.024$, 0.347 & $-0.012$, 0.345 \\
    };
    \end{tikzpicture}
    \caption{All the results of training the three astrophysical parameter models $A_\text{SN1}$, $A_\text{SN2}$, $A_\text{AGN}$ on every combination of node features used, as well as the additional main branch and flattened tree tests (columns 6 and 7, respectively). Each entry in the table shows the calculated coefficient of determination and the root mean squared error, respectively.}
    \label{tab:allparamresults}
\end{figure}

\end{document}